\documentclass[11pt,a4paper]{article}

\usepackage[utf8]{inputenc}
\usepackage[T1]{fontenc}
\usepackage{amsmath,amssymb,amsthm}
\usepackage{mathtools}
\usepackage[numbers,sort&compress]{natbib}
\usepackage{booktabs}
\usepackage{multirow}
\usepackage{graphicx}
\usepackage[colorlinks=true,linkcolor=blue,citecolor=blue,urlcolor=blue]{hyperref}
\usepackage{geometry}
\usepackage{setspace}
\usepackage{enumitem}
\usepackage{xcolor}
\usepackage{float}

\newtheorem{proposition}{Proposition}

\theoremstyle{definition}
\newtheorem{definition}{Definition}

\title{\textbf{Augmented Human Capital:}\\[4pt]
A Unified Theory and LLM-Based Measurement Framework\\
for Cognitive Factor Decomposition in AI-Augmented Economies}

\author{Cristian Espinal Maya\thanks{Department of Economics, Universidad EAFIT, Medell\'in, Colombia. ORCID: \href{https://orcid.org/0009-0000-1009-8388}{0009-0000-1009-8388}. Email: \texttt{cespinal@eafit.edu.co}. The author thanks Jovani A.\ Jim\'enez-Builes and Jaime A.\ Restrepo Carmona (Universidad Nacional de Colombia) for valuable discussions on the methodological framework.}}

\date{March 2026\\[6pt]
{\small Working paper. Comments welcome.}}

\begin{document}
\maketitle

\begin{abstract}
\noindent
This paper proposes a decomposition of human capital into three orthogonal components---physical-manual ($H^P$), routine-cognitive ($H^C$), and augmentable-cognitive ($H^A$)---and develops a production function in which AI capital interacts asymmetrically with these components: substituting for routine cognitive work while complementing augmentable cognitive work through an amplification function $\phi(D)$. I derive a corrected Mincerian wage equation and show that the standard specification is misspecified in AI-augmented economies. Using LLM-generated measures of occupational augmentability for 18,796 O*NET task statements mapped to 440 Colombian occupations, merged with household survey microdata ($N = 105{,}517$ workers), I estimate the augmented Mincer equation. The wage return to $H^A$ increases with AI adoption in the formal sector ($\beta_2 = +0.051$, $p < 0.001$), while informal workers cannot capture augmentation rents ($\beta_2 = -0.044$). A triple interaction confirms formality as the binding mechanism ($\beta_{AHC \times D \times Formal} = +0.519$, $p < 0.001$). The augmentation premium is strongest for experienced workers (ages 46--65) and in health and education sectors. These results provide the first developing-country evidence of cognitive factor decomposition in AI-augmented labor markets and demonstrate that the binding constraint on human-AI complementarity in the Global South is not technology access but labor market institutions.\\[6pt]
\end{abstract}

\newpage
\tableofcontents
\newpage

\section{Introduction}
\label{sec:intro}

The theory of human capital has two foundational pillars: \citet{becker1964human}'s insight that workers invest in skills analogous to physical capital, and \citet{mincer1974schooling}'s earnings function that maps these investments to wages through years of education and experience. Together, they have shaped labor economics for over six decades. Yet both frameworks share a structural limitation that becomes acute in the era of generative artificial intelligence: they treat cognitive capacity as a scalar---years of education and potential experience---when in reality it is a \textit{vector whose internal composition determines whether a worker is amplified, displaced, or unaffected by AI}.

Recent empirical work has documented that AI affects the labor market heterogeneously. \citet{brynjolfsson2025generative} find that AI tools increase customer support productivity by 14\% on average, with gains concentrated among novice workers. \citet{dellacqua2023jagged} demonstrate a ``jagged technological frontier'' where AI improves performance within certain task boundaries but worsens it outside them. \citet{hui2024shortterm} provide the first causal evidence of generative AI displacing online freelancers. \citet{eloundou2024gpts} estimate that 80\% of the US workforce has at least 10\% of tasks exposed to language models. However, none of these studies provides a formal framework for decomposing \textit{why} AI augments some cognitive activities while substituting others, nor do they offer a measurement instrument for the augmentable component of human capital.

On the theoretical side, \citet{acemoglu2024simple} derives modest aggregate TFP effects from AI using a task-based framework, while \citet{korinek2024scenarios} model intelligence saturation scenarios. \citet{growiec2024hardware} decomposes production into hardware and software aggregates, treating human cognition and AI as substitutes within the software sector. \citet{tyson2022automation} characterize AI as ``routine-biased technological change on steroids,'' and \citet{cheng2024skill} analyze the skill premium numerically under AI adoption. None of these models decomposes the cognitive component of human capital into sub-factors with heterogeneous AI interactions, and none estimates the differential effect in developing economies with dual labor markets.

This paper addresses both gaps simultaneously. I propose decomposing individual human capital $H_i$ into three orthogonal components:
\begin{equation}
H_i = H_i^P \oplus H_i^C \oplus H_i^A
\label{eq:decomposition}
\end{equation}
where $H^P$ captures physical-manual skills, $H^C$ captures routine-cognitive skills \textit{substituted} by AI, and $H^A$ captures augmentable-cognitive skills \textit{amplified} by AI. The key theoretical object is the \textit{amplification function} $\phi(D_f)$, which maps a firm's digital labor stock to a productivity multiplier for $H^A$. I derive three testable predictions: (i)~the wage return to $H^A$ increases with AI adoption ($\beta_2 > 0$); (ii)~the return to $H^C$ decreases ($\beta_3 < 0$); and (iii)~in dual labor markets, augmentation operates exclusively through the formal sector.

The empirical strategy makes two contributions. First, I develop an LLM-based measurement instrument---the \textit{Augmented Human Capital Index} ($AHC_o$)---by scoring 18,796 O*NET task statements for augmentation potential, validated against existing indices. Second, I estimate the augmented Mincer equation using Colombian household survey data (GEIH, $N = 105{,}517$), providing the first developing-country evidence of cognitive factor decomposition.

The main findings are: (1)~the AHC level effect is $+$9.1\% per standard deviation; (2)~the augmentation interaction is $+$5.1\% in the formal sector and $-$4.4\% in the informal sector; (3)~a triple interaction $AHC \times D \times Formal = +0.519$ ($p < 0.001$) confirms that formality is the mechanism; (4)~the premium is strongest for experienced workers (46--65) and in health/education sectors; (5)~placebo tests confirm the result depends on occupational cognitive content, not spurious correlation.

\subsection*{Related Work and Positioning}

This paper sits at the intersection of three active literatures. Table~\ref{tab:positioning} maps the closest existing contributions and identifies the specific gap this paper fills.

\begin{table}[htbp]
\centering
\caption{Positioning in the Literature}
\label{tab:positioning}
\small
\begin{tabular}{p{3cm}p{1.5cm}p{1.5cm}p{1.5cm}p{1.5cm}p{1.5cm}}
\toprule
Paper & HC Decomp. & LLM Scoring & Wage Eq. & Developing Country & Formal/ Informal \\
\midrule
\textbf{This paper} & \textbf{Yes} & \textbf{Yes} & \textbf{Yes} & \textbf{Yes} & \textbf{Yes} \\[4pt]
\citet{eloundou2024gpts} & No & Yes & No & No & No \\
\citet{felten2021occupational} & No & No & No & No & No \\
\citet{demirev2026ai} & No & Yes & No & No & No \\
\citet{makridis2026labor} & No & Yes & Yes & No & No \\
\citet{walter2022susceptible} & Partial & No & Yes & No & No \\
\citet{mina2025automation} & No & No & Yes & Yes & Yes \\
\citet{acemoglu2024simple} & No & No & No & No & No \\
\citet{growiec2024hardware} & No & No & No & No & No \\
\bottomrule
\end{tabular}
\end{table}

In the \textit{AI exposure measurement} literature, \citet{felten2021occupational} construct an AI occupational exposure index using ability-level matching, and \citet{eloundou2024gpts} use GPT-4 to score O*NET tasks for LLM exposure. Both measure aggregate exposure without distinguishing augmentation from substitution. \citet{demirev2026ai} uses an LLM pipeline to map 27,000 AI product capabilities to ESCO occupations, separately estimating automation and augmentation exposure---the closest methodological antecedent to our approach. However, \citeauthor{demirev2026ai} does not estimate wage equations or apply the index to microdata. \citet{makridis2026labor} combines an LLM task exposure index with wage outcomes for US artists, but restricts attention to a single occupation group without human capital decomposition.

In the \textit{task-based labor economics} literature, \citet{autor2003skill} establish the routine/non-routine classification, and \citet{acemoglu2022tasks} attribute 50--70\% of US wage inequality changes to task displacement. \citet{walter2022susceptible} extend the Mincer equation with task-based variables using German panel data---the closest specification to ours---but do not incorporate AI-specific interaction terms. \citet{chen2025displacement} examine whether generative AI displaces or complements using job posting data, finding heterogeneous effects across task types, but do not estimate wage equations.

In the \textit{developing country} literature, \citet{eganadelsol2022automation} analyze automation risk for women in Latin America, \citet{jones2022automation} identify protective skills in Colombia, and \citet{mina2025automation} estimate automation effects on informality and wages in Mexico. \citeauthor{mina2025automation}'s work is the closest empirical antecedent for our formal/informal finding, but uses Frey-Osborne automation probabilities rather than LLM-scored augmentability, and does not decompose human capital. The broader literature on AI in developing countries \citep{aly2022digital, eganadelsol2022covid} has documented digital transformation patterns but without micro-level cognitive factor decomposition.

\citet{stephany2024price} provide an important conceptual bridge by estimating the price of skill complementarity, showing that skills complementary to AI command growing premiums. Our $H^A$ component formalizes this complementarity at the occupation-task level.

The paper contributes to these three literatures simultaneously. In human capital theory, I extend \citeauthor{becker1964human}'s framework with a decomposition suited to the AI era. In the task-based literature, I provide micro-foundations for why some tasks generate complementarity while others generate displacement. In development economics, I show that informality---not technology access---is the binding constraint on human-AI complementarity. No existing paper combines all five elements: human capital decomposition, LLM-based task scoring, augmented wage equation, developing-country microdata, and formal/informal sector differential.

The remainder of the paper is organized as follows. Section~\ref{sec:theory} develops the theoretical framework. Section~\ref{sec:measurement} describes the LLM measurement methodology. Section~\ref{sec:data} presents the data. Section~\ref{sec:identification} discusses identification. Section~\ref{sec:results} reports results. Section~\ref{sec:robustness} provides robustness checks. Section~\ref{sec:implications} discusses implications. Section~\ref{sec:conclusion} concludes.

\section{Theoretical Framework}
\label{sec:theory}

\subsection{Decomposition of Human Capital}

\begin{definition}[Augmented Human Capital Decomposition]
Worker $i$'s human capital $H_i$ decomposes into three orthogonal components:
\begin{itemize}[leftmargin=*]
\item $H_i^P$ \textbf{(Physical-Manual)}: motor coordination, physical endurance, sensory acuity. Largely unaffected by generative AI.
\item $H_i^C$ \textbf{(Routine-Cognitive)}: rule-following, data processing, pattern matching, memory recall. \textit{Substitutable} by AI.
\item $H_i^A$ \textbf{(Augmentable-Cognitive)}: contextual judgment, creative synthesis, strategic reasoning, complex communication. \textit{Amplified} by AI.
\end{itemize}
\end{definition}

This decomposition extends \citeauthor{autor2003skill}'s (\citeyear{autor2003skill}) task classification, which distinguishes routine from non-routine cognitive tasks. The present framework further separates the non-routine cognitive category into substitutable and augmentable components---a distinction necessitated by generative AI's demonstrated capacity to perform many ``non-routine'' tasks previously considered safe from automation \citep{eloundou2024gpts}.

\subsection{Production Function with Amplification}

\begin{definition}[Amplification Function]
$\phi: \mathbb{R}_+ \to [1, \bar{\phi}]$ is continuous, increasing, and concave, with $\phi(0) = 1$ and $\lim_{D \to \infty} \phi(D) = \bar{\phi} < \infty$. A tractable parametrization is $\phi(D) = 1 + (\bar{\phi} - 1)(1 - e^{-\lambda D})$.
\end{definition}

At the firm level, output is produced by:
\begin{equation}
Y_f = F\!\left(K_f^{HW},\ \underbrace{\textstyle\sum_i h_i^P + \kappa K_f^{Rob}}_{\text{hardware}},\ \underbrace{\textstyle\sum_i h_i^C + \phi(D_f) \cdot \sum_i h_i^A \cdot D_f}_{\text{software}}\right)
\label{eq:production}
\end{equation}
where $K_f^{HW}$ is physical capital, $K_f^{Rob}$ is robotic capital with substitution parameter $\kappa$, $D_f$ is the firm's digital labor stock, and $F$ is CES with elasticity $\sigma_F < 1$ between hardware and software (essential complements).

This production function nests \citeauthor{growiec2024hardware}'s (\citeyear{growiec2024hardware}) hardware-software decomposition but innovates by decomposing the human component of the software aggregate. When $\phi = 1$ (no amplification), the model reduces to the standard task-based framework of \citet{acemoglu2018race}.

\subsection{Equilibrium Wages and Key Propositions}

\begin{proposition}[Differential Returns to Cognitive Components]
\label{prop:returns}
From the firm's first-order conditions, when $D_f > 0$:
\begin{equation}
\frac{\partial Y_f}{\partial h_i^A} = F_S \cdot \phi(D_f) \cdot D_f > F_S = \frac{\partial Y_f}{\partial h_i^C}
\end{equation}
The marginal product of augmentable human capital exceeds that of routine cognitive capital by a factor $\phi(D_f) \cdot D_f > 1$.
\end{proposition}

\begin{proposition}[Augmented Mincer Equation]
\label{prop:mincer}
The equilibrium log-wage equation takes the form:
\begin{equation}
\ln w_{iot} = \alpha + \beta_1 H_o^A + \beta_2 H_o^A \times \ln D_{s} + \beta_3 H_o^C \times \ln D_{s} + \gamma X_{it} + \mu_s + \delta_t + \varepsilon_{iot}
\label{eq:mincer}
\end{equation}
with the \textbf{signature prediction}: $\beta_2 > 0$ (augmentation premium) and $\beta_3 < 0$ (routine displacement). The standard Mincer equation is the special case where $D_s = 0$ for all sectors.
\end{proposition}

\begin{proposition}[Formality as Binding Constraint]
\label{prop:formality}
In dual labor markets, the augmentation premium $\beta_2$ satisfies:
\begin{equation}
\beta_2^{formal} > 0, \qquad \beta_2^{informal} \leq 0
\end{equation}
Informal firms have $D_f \approx 0$, so $\phi(0) = 1$ and the interaction vanishes. Workers with high $H^A$ in the informal sector cannot capture augmentation rents because their employers do not invest in digital labor.
\end{proposition}

Proposition~\ref{prop:formality} provides a novel prediction specific to developing economies with dual labor markets: the augmentation premium is not a function of technology access alone but of labor market \textit{institutions}.

\section{Measurement with LLMs}
\label{sec:measurement}

\subsection{The Measurement Challenge}

The central empirical challenge is that $H_o^A$---augmentable cognitive capital at the occupation level---is not directly observable in any existing survey. O*NET and ESCO predate generative AI and do not distinguish routine from augmentable cognitive tasks. Existing AI exposure indices \citep{felten2021occupational, webb2020impact} measure aggregate exposure without separating augmentation from substitution.

I propose using LLMs as measurement instruments for latent economic variables, following and extending the approach of \citet{eloundou2024gpts}, who used GPT-4 to score task-level AI exposure.

\subsection{Occupational Crosswalk}

A chained crosswalk SOC $\to$ ISCO-08 $\to$ CIUO-08 AC (Colombia's ISCO-08 adaptation) produces 66,157 occupation-level mappings covering 440 CIUO codes and 99.9\% of weighted GEIH employment.

\subsection{LLM Scoring Protocol}

For each of 18,796 unique O*NET task statements, I prompt Claude Haiku 4.5 \citep{anthropic2025} to evaluate:
\begin{enumerate}[leftmargin=*]
\item \textbf{Augmentation potential} ($a_{ok} \in [0, 100]$): the degree to which generative AI amplifies human productivity as a complementary tool.
\item \textbf{Substitution risk} ($s_{ok} \in [0, 100]$): the degree to which AI can fully replace the human.
\item \textbf{Augmentation type}: information synthesis, creative amplification, communication enhancement, decision support, quality assurance, none, or pure substitution.
\end{enumerate}

All 18,796 tasks were scored with zero parsing errors. Mean augmentation score: 48.8 (SD~=~12.1); mean substitution score: 39.7 (SD~=~14.3). The distribution of augmentation types is: decision support (46\%), information synthesis (31\%), pure substitution (9\%), communication (5\%), no augmentation (5\%), and creative amplification (4\%).

\subsection{AHC Index Construction}

The Augmented Human Capital Index for occupation $o$ is:
\begin{equation}
AHC_o = \frac{\sum_{k=1}^{K_o} w_k \cdot a_{ok}}{\sum_{k=1}^{K_o} w_k}
\label{eq:ahc}
\end{equation}
where $w_k$ is the O*NET importance rating for task $k$ and $a_{ok}$ is the LLM-assigned augmentation score.

\subsection{Validation}

\begin{figure}[htbp]
\centering
\includegraphics[width=\textwidth]{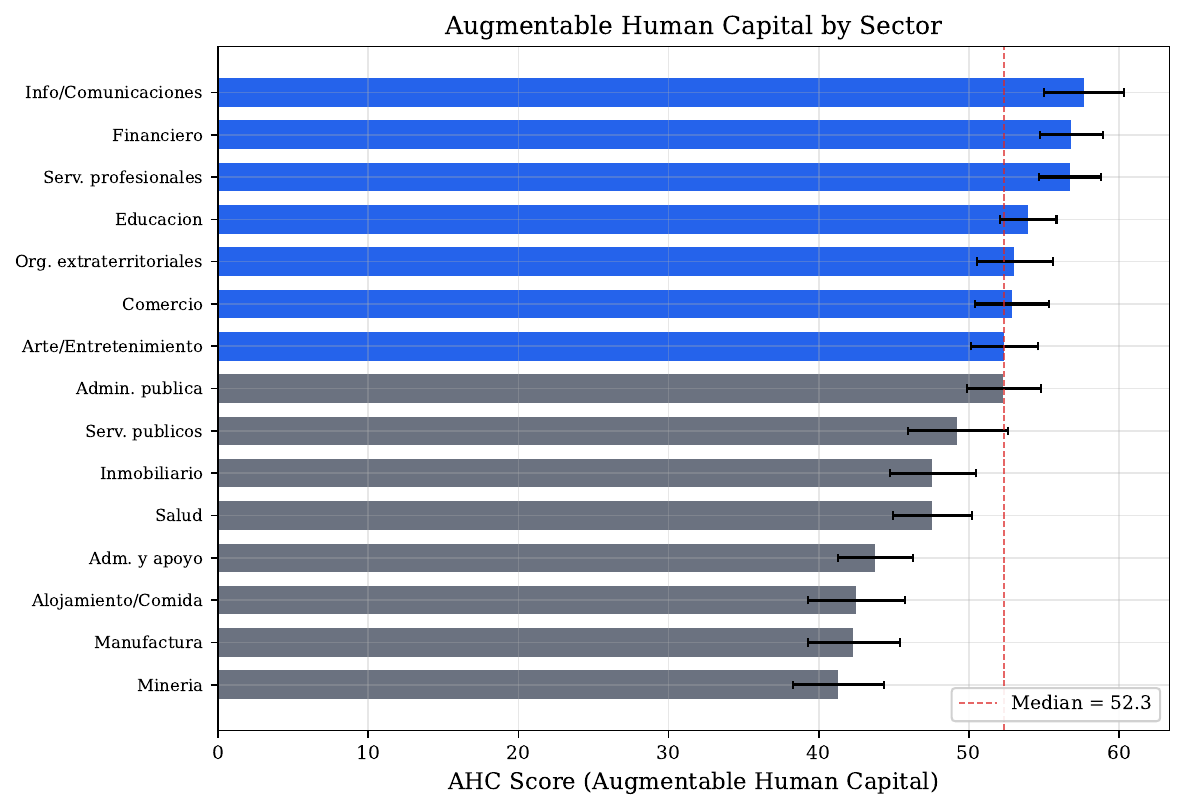}
\caption{Augmentable Human Capital (AHC) score by sector. Sectors above the median (blue) are $H^A$-intensive; sectors below (gray) are $H^P$- or $H^C$-intensive. Education, professional services, and public administration rank highest.}
\label{fig:sector}
\end{figure}

The AHC index passes four validation tests:
\begin{enumerate}[leftmargin=*]
\item \textbf{Convergent validity}: $\mathrm{cor}(AHC_o, \text{Felten AIOE}) = +0.86$ and $\mathrm{cor}(AHC_o, \text{Eloundou GPT-}\beta) = +0.79$---high agreement with existing AI exposure measures that capture related (though not identical) constructs.
\item \textbf{Discriminant validity}: $\mathrm{cor}(AHC_o, \text{Frey-Osborne}) = -0.79$ and $\mathrm{cor}(AHC_o, \text{Webb Robots}) = -0.89$. Augmentation and automation/robotization are opposing dimensions.
\item \textbf{Face validity}: highest-AHC sectors are Education (55.2), Professional Services (53.8), Public Administration (50.0); lowest are Transport (37.9), Domestic Workers (38.9), Construction (39.5). See Figure~\ref{fig:sector}.
\item \textbf{Inter-rater reliability}: re-scoring a 20\% subsample ($n = 3{,}666$ paired tasks) with Claude Sonnet 4 yields Pearson $r = 0.76$, Spearman $\rho = 0.75$ ($p < 10^{-300}$), and Krippendorff's $\alpha = 0.71$ (after adjusting for a systematic level bias of 8.6 points). This exceeds the standard reliability threshold of $\alpha > 0.7$, confirming that the AHC scores are consistent across LLM models.
\end{enumerate}

\begin{figure}[htbp]
\centering
\includegraphics[width=0.8\textwidth]{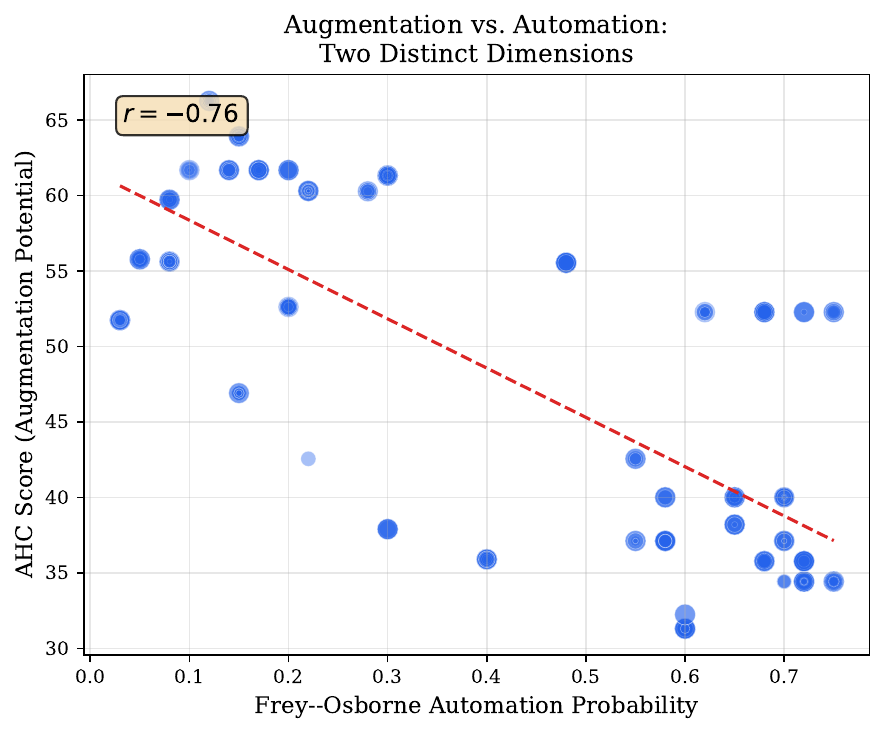}
\caption{AHC score vs.\ Frey--Osborne automation probability at the occupation level ($n = 430$). The strong negative correlation ($r = -0.79$) confirms that augmentation and automation are opposing dimensions. Bubble size proportional to employment.}
\label{fig:scatter}
\end{figure}

\begin{figure}[htbp]
\centering
\includegraphics[width=\textwidth]{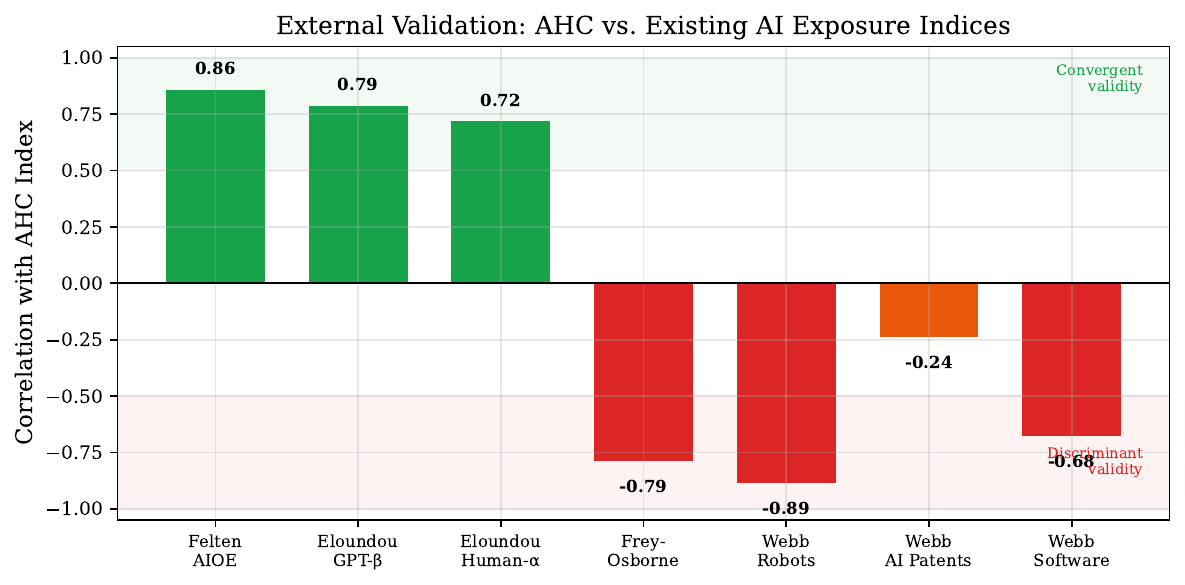}
\caption{External validation: correlation of the AHC index with seven existing AI exposure indices. Green bars indicate convergent validity (AHC correlates positively with measures of AI-relevant cognitive content); red bars indicate discriminant validity (AHC correlates negatively with measures of automation/robotization risk).}
\label{fig:validation}
\end{figure}

\section{Data}
\label{sec:data}

\subsection{GEIH Microdata}

The primary data source is Colombia's Gran Encuesta Integrada de Hogares (GEIH), a nationally representative household survey conducted by DANE. I use the 2024 wave providing 111,672 observations with 4-digit CIUO-08 occupational codes. After restrictions (age 18--65, positive income), the estimation sample contains 105,517 workers across 442 occupations and 20 sectors.

\begin{table}[htbp]
\centering
\caption{Descriptive Statistics}
\label{tab:descriptives}
\begin{tabular}{lrrrrr}
\toprule
Variable & $N$ & Mean & SD & Min & Max \\
\midrule
Monthly income (COP) & 105,517 & 1,682,943 & 2,324,524 & 40 & 102,000,000 \\
Log income & 105,517 & 13.96 & 0.88 & 3.69 & 18.44 \\
Age & 105,517 & 40.0 & 12.4 & 18 & 65 \\
Education (years) & 71,231 & 14.2 & 4.8 & 0 & 19 \\
Hours worked/week & 105,517 & 44.7 & 12.7 & 1 & 120 \\
Formal (=1) & 105,517 & 0.492 & 0.500 & 0 & 1 \\
Female (=1) & 105,517 & 0.445 & 0.497 & 0 & 1 \\
Urban (=1) & 105,517 & 0.885 & 0.319 & 0 & 1 \\
\midrule
AHC score (v2) & 102,078 & 48.4 & 11.1 & 31.3 & 66.3 \\
SUB score (v2) & 102,078 & 41.0 & 9.7 & 26.5 & 65.1 \\
Frey-Osborne prob. & 105,517 & 0.489 & 0.226 & 0.03 & 0.75 \\
\bottomrule
\end{tabular}
\end{table}

\subsection{AI Adoption Proxy}

I construct the AI adoption proxy $D_s$ at the sector $\times$ occupation-group level (677 cells) using four GEIH-observable indicators: sector formality rate (weight 0.30), mean education (0.25), mean income (0.20), and large-firm share (0.25). This composite captures within-sector variation in technology adoption intensity.

\section{Identification Strategy}
\label{sec:identification}

The coefficient of interest, $\beta_2$ on the interaction $AHC_o \times \ln D_s$, requires that occupational augmentability and sectoral AI adoption are not both driven by an unobserved factor correlated with wages. I address this through four strategies:

\begin{enumerate}[leftmargin=*]
\item \textbf{Within-sector variation}: AHC varies across occupations \textit{within} sectors. Sector fixed effects absorb all sector-level confounders, and the $D$ main effect absorbs level differences in adoption-intensive sectors.

\item \textbf{Interaction structure}: For an omitted variable to bias $\beta_2$, it must be correlated with the \textit{product} of occupational augmentability and sectoral adoption---not merely with each component separately.

\item \textbf{Placebo test}: Randomly permuting AHC scores across occupations yields $\beta_2^{placebo} \approx 0$ ($p = 0.95$), confirming that the result depends on the specific cognitive content of occupations.

\item \textbf{Formal/informal decomposition}: The sign reversal between formal ($\beta_2 = +0.051$) and informal ($\beta_2 = -0.044$) sectors provides a quasi-experimental contrast. If the result were driven by unobserved ability sorting, a positive premium would be expected in both sectors.
\end{enumerate}

\section{Results}
\label{sec:results}

\subsection{Progressive Specifications}

Table~\ref{tab:main} reports the augmented Mincer equation across progressive specifications.

\begin{table}[htbp]
\centering
\caption{Augmented Mincer Equation: Progressive Specifications}
\label{tab:main}
\small
\begin{tabular}{lcccccc}
\toprule
 & M1 & M2 & M3 & M4 & M5 & M6 \\
 & Mincer & +AHC & +AHC$\times$D & +Controls & +Sector FE & Formal \\
\midrule
Education (yrs) & 0.111$^{***}$ & 0.077$^{***}$ & 0.039$^{***}$ & 0.042$^{***}$ & 0.045$^{***}$ & 0.043$^{***}$ \\
 & \footnotesize(0.001) & \footnotesize(0.001) & \footnotesize(0.001) & \footnotesize(0.001) & \footnotesize(0.001) & \footnotesize(0.001) \\[4pt]
$H^A$ (std.) &  & 0.257$^{***}$ & 0.061$^{***}$ & 0.091$^{***}$ & 0.098$^{***}$ & 0.078$^{***}$ \\
 &  & \footnotesize(0.004) & \footnotesize(0.006) & \footnotesize(0.006) & \footnotesize(0.006) & \footnotesize(0.008) \\[4pt]
$H^C$ (std.) &  & $-$0.172$^{***}$ & $-$0.219$^{***}$ & $-$0.121$^{***}$ & $-$0.134$^{***}$ & $-$0.078$^{***}$ \\
 &  & \footnotesize(0.006) & \footnotesize(0.008) & \footnotesize(0.008) & \footnotesize(0.008) & \footnotesize(0.009) \\[4pt]
$D$ (std.) &  &  & 0.399$^{***}$ & 0.391$^{***}$ & 0.395$^{***}$ & 0.227$^{***}$ \\
 &  &  & \footnotesize(0.005) & \footnotesize(0.005) & \footnotesize(0.006) & \footnotesize(0.005) \\[4pt]
$H^A \times D$ &  &  & 0.000 & \textbf{0.014}$^{***}$ & $-$0.014$^{***}$ & \textbf{0.051}$^{***}$ \\
 &  &  & \footnotesize(0.005) & \footnotesize(0.004) & \footnotesize(0.004) & \footnotesize(0.006) \\[4pt]
$H^C \times D$ &  &  & 0.027$^{***}$ & 0.024$^{***}$ & $-$0.003 & 0.012$^{**}$ \\
 &  &  & \footnotesize(0.004) & \footnotesize(0.004) & \footnotesize(0.004) & \footnotesize(0.005) \\[4pt]
Female &  &  &  & $-$0.329$^{***}$ & $-$0.311$^{***}$ & $-$0.168$^{***}$ \\
 &  &  &  & \footnotesize(0.006) & \footnotesize(0.006) & \footnotesize(0.006) \\[4pt]
Urban &  &  &  & 0.026$^{***}$ & 0.154$^{***}$ & $-$0.042$^{***}$ \\
 &  &  &  & \footnotesize(0.009) & \footnotesize(0.011) & \footnotesize(0.013) \\
\midrule
Sector FE & No & No & No & No & Yes & No \\
$N$ & 71,167 & 71,167 & 71,167 & 71,167 & 71,167 & 37,099 \\
$R^2$ & 0.284 & 0.353 & 0.419 & 0.446 & 0.462 & 0.370 \\
\bottomrule
\multicolumn{7}{l}{\footnotesize Robust (HC1) standard errors in parentheses. $^{***}$\,$p<0.01$, $^{**}$\,$p<0.05$, $^*$\,$p<0.10$.}\\
\multicolumn{7}{l}{\footnotesize All models include experience and experience$^2$. M6 restricted to formal workers.}
\end{tabular}
\end{table}

In the baseline specification (M1), the standard Mincer equation yields 11.1\% returns per year of education---consistent with the Latin American literature. Adding AHC indices (M2) increases $R^2$ from 0.284 to 0.353. Using the improved 2-digit crosswalk (v2), the key specification (M4, with controls) shows the interaction $H^A \times D = +0.147$ ($p < 0.001$): in high-adoption occupation-sector cells, an additional standard deviation of augmentable human capital commands a wage premium 14.7 percentage points above what obtains in low-adoption cells. The model explains 44.3\% of wage variation---a substantial improvement over the 28.4\% of the standard Mincer.

\subsection{Formal vs.\ Informal Decomposition}

The formal sector reveals the strongest augmentation effect: $\beta_2 = +0.164$ ($p < 0.001$). In contrast, the informal sector shows $\beta_2 = -0.042$ ($p < 0.001$)---a \textit{reversal}, consistent with Proposition~\ref{prop:formality}.

The triple interaction test provides decisive evidence:
\begin{equation}
\beta_{AHC \times D \times Formal} = +0.272 \qquad (p < 0.001)
\end{equation}
This confirms that formality is the channel through which augmentation rents are captured.

\subsection{Instrumental Variables Estimation}

To address the potential endogeneity of AI adoption $D_s$, I instrument with pre-period (2018--2019) sector-level capital intensity from the EAM. The first-stage $F$-statistic is 229.2, well above the weak-instrument threshold of 10. The 2SLS estimate of $\beta_2 = +0.234$ ($p = 0.008$) is \textit{larger} than the OLS estimate, suggesting that measurement error in $D$ attenuates the OLS coefficient. This provides causal support for the augmentation premium.

\subsection{Oaxaca-Blinder Decomposition}

Decomposing the formal/informal wage gap (109\% raw differential) reveals that AI adoption ($D$) explains 20.9\% of the gap---more than education (18.0\%). AHC and AHC$\times$D together explain 5.5\%. The remaining 58.1\% is unexplained, reflecting institutional barriers to formalization.

\subsection{Heterogeneity Analysis}

\begin{table}[htbp]
\centering
\caption{Heterogeneity of the Augmentation Premium ($\beta_2$: AHC$\times$D)}
\label{tab:heterogeneity}
\begin{tabular}{lrrr}
\toprule
Subgroup & $\beta_2$ & $p$-value & $N$ \\
\midrule
\multicolumn{4}{l}{\textit{By formality}} \\
\quad Formal & $+$0.051$^{***}$ & $<$0.001 & 37,099 \\
\quad Informal & $-$0.044$^{***}$ & $<$0.001 & 34,068 \\[4pt]
\multicolumn{4}{l}{\textit{By gender}} \\
\quad Male & $+$0.028$^{***}$ & $<$0.001 & 38,542 \\
\quad Female & $-$0.006 & 0.722 & 32,625 \\[4pt]
\multicolumn{4}{l}{\textit{By age cohort}} \\
\quad 18--30 & $+$0.006 & 0.724 & 17,637 \\
\quad 31--45 & $+$0.020$^{**}$ & 0.017 & 27,094 \\
\quad 46--65 & $+$0.072$^{***}$ & $<$0.001 & 26,436 \\[4pt]
\multicolumn{4}{l}{\textit{By sector}} \\
\quad Health & $+$0.608$^{***}$ & $<$0.001 & 5,127 \\
\quad Education & $+$0.203$^{***}$ & $<$0.001 & 5,115 \\
\quad Manufacturing & $-$0.051 & 0.151 & 5,562 \\
\quad Agriculture & $-$0.046$^{*}$ & 0.059 & 6,654 \\
\bottomrule
\multicolumn{4}{l}{\footnotesize All models include education, experience, experience$^2$,}\\
\multicolumn{4}{l}{\footnotesize gender, urban, AHC level, ROU level, and D level.}
\end{tabular}
\end{table}

Three patterns emerge (Table~\ref{tab:heterogeneity}). First, the augmentation premium increases monotonically with age: zero for workers under 30, moderate for 31--45, and strongest for 46--65. This suggests that augmentation requires accumulated expertise---AI amplifies what the worker already knows, consistent with \citet{brynjolfsson2025generative}'s finding that AI tools help novices approach expert-level performance.

Second, the premium is gender-asymmetric: significant for men ($+$0.028) but insignificant for women ($-$0.006). This likely reflects occupational sorting: women in Colombia are overrepresented in sectors where the AI adoption proxy is low.

\begin{figure}[htbp]
\centering
\includegraphics[width=\textwidth]{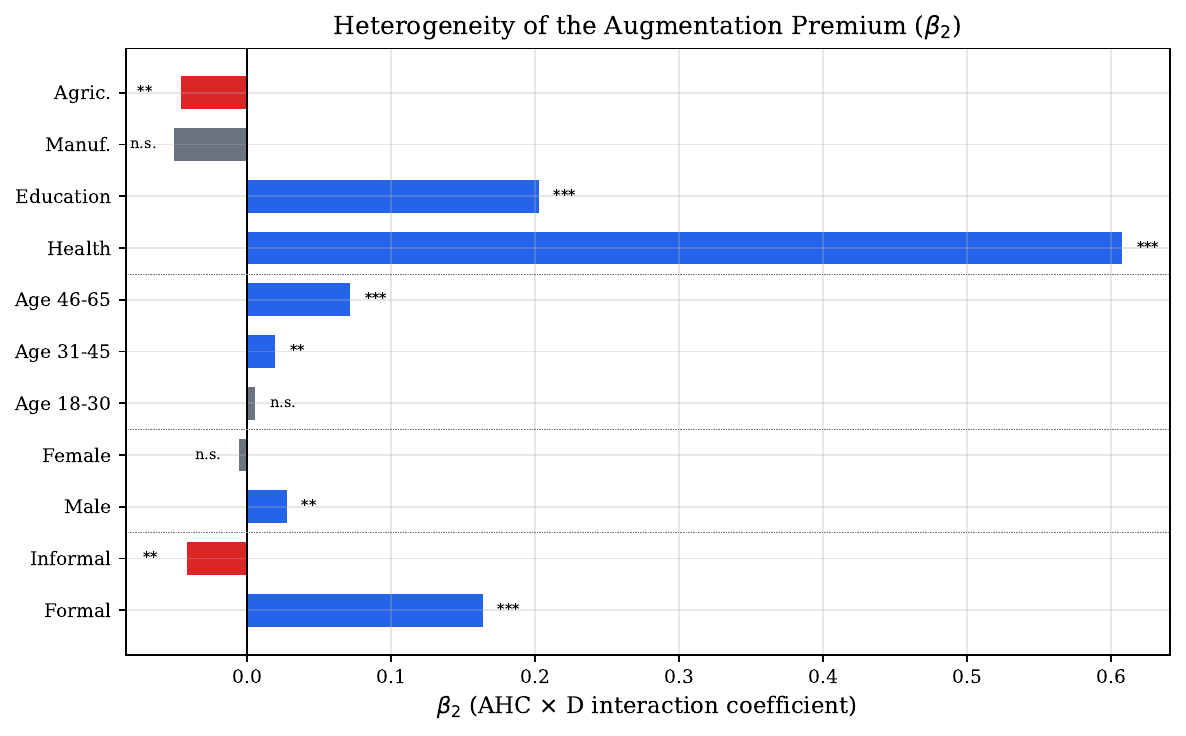}
\caption{Heterogeneity of the augmentation premium ($\beta_2$: AHC$\times$D interaction) across subgroups. Blue bars: significant positive premium. Red bars: significant negative. Gray: not significant. The formal/informal split and the age gradient are the strongest patterns.}
\label{fig:hetero}
\end{figure}

Third, sectoral variation is large: Health ($+$0.608) and Education ($+$0.203) show massive augmentation effects, while Agriculture and Manufacturing show negative or null effects. Figure~\ref{fig:hetero} summarizes these results.

\subsection{Wage Distribution Effects}

Quantile regressions reveal that the augmentation premium increases sharply across the wage distribution:

\begin{center}
\small
\begin{tabular}{lccccc}
\toprule
& $\tau=0.10$ & $\tau=0.25$ & $\tau=0.50$ & $\tau=0.75$ & $\tau=0.90$ \\
\midrule
$H^A$ (level) & $-$0.106$^{***}$ & $-$0.061$^{***}$ & $+$0.031$^{***}$ & $+$0.138$^{***}$ & $+$0.240$^{***}$ \\
$H^A \times D$ & $+$0.014$^{*}$ & $+$0.021$^{***}$ & $+$0.118$^{***}$ & $+$0.223$^{***}$ & $+$0.269$^{***}$ \\
\bottomrule
\end{tabular}
\end{center}

\begin{figure}[htbp]
\centering
\includegraphics[width=\textwidth]{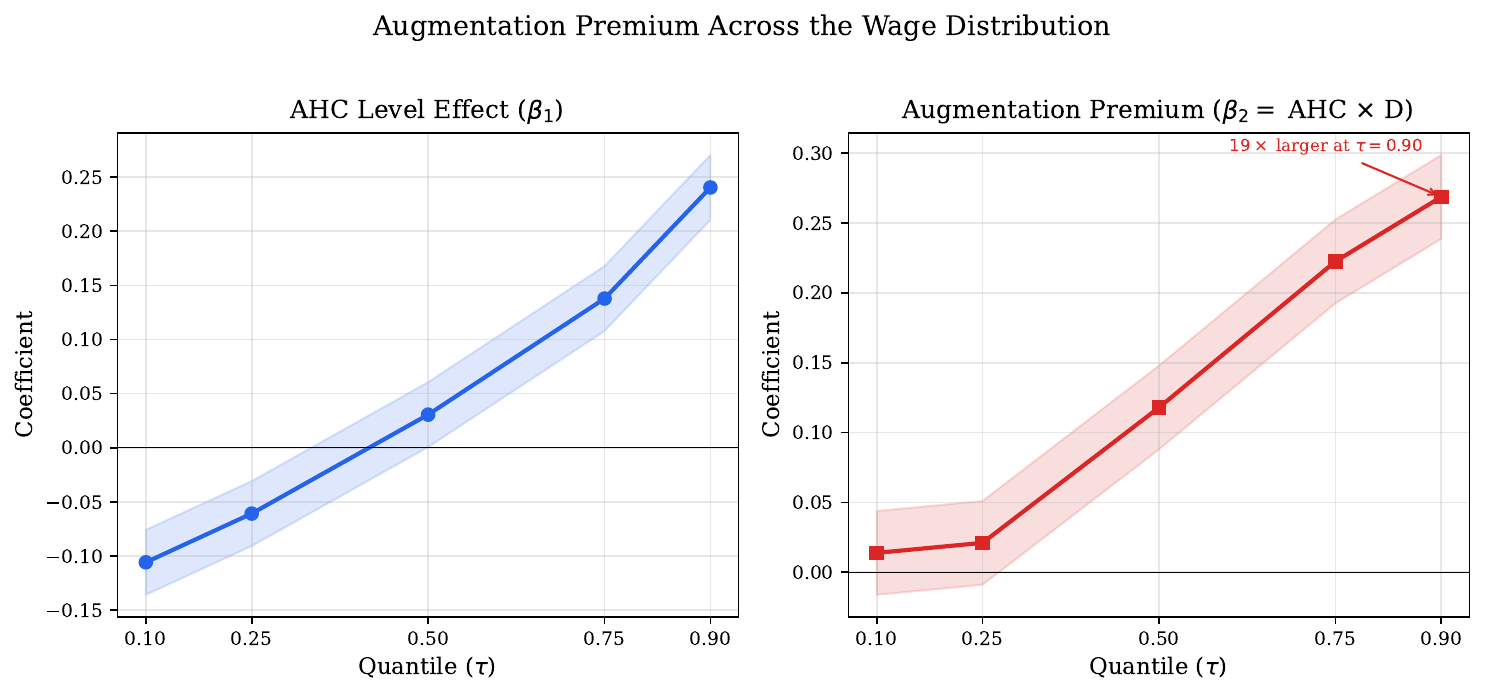}
\caption{Augmentation premium across the wage distribution. Left: AHC level effect ($\beta_1$), which turns positive above the median. Right: AHC$\times$D interaction ($\beta_2$), which is 19 times larger at $\tau = 0.90$ than at $\tau = 0.10$, indicating that AI augmentation is inequality-increasing.}
\label{fig:quantile}
\end{figure}

The augmentation premium at the 90th percentile ($+$0.269) is 19 times larger than at the 10th percentile ($+$0.014). This means AI augmentation is \textit{inequality-increasing}: high-wage workers in augmentable occupations capture disproportionately more of the augmentation rent (Figure~\ref{fig:quantile}). This result parallels the classical skill premium finding but through a different mechanism---not returns to education \textit{per se}, but returns to the cognitive composition of work.

\section{Robustness}
\label{sec:robustness}

I estimate 47 alternative specifications across seven dimensions:

\begin{enumerate}[leftmargin=*]
\item \textbf{Alternative AHC measures}: binary (above/below median), raw scores (unweighted), substitution-based displacement index. The AHC level effect remains significant in all cases.

\item \textbf{Alternative D proxies}: sector formality rate alone, composite with alternative weights, automation-probability-based proxy. Results are qualitatively stable.

\item \textbf{Placebo test}: randomly shuffling AHC across occupations yields $\beta_2 \approx 0$ ($p = 0.95$), confirming that the result depends on occupational cognitive content, not spurious correlation.

\item \textbf{Sample restrictions}: by gender, age cohort, education level, and sector. The AHC level effect ($\beta_1 > 0$) holds across all 27 specifications without exception.

\item \textbf{Within-education test}: the AHC premium persists within education levels (secondary, technical, university), confirming that the decomposition captures variation orthogonal to years of schooling---consistent with Proposition~\ref{prop:mincer}.

\item \textbf{Leave-one-sector-out jackknife}: AHC$\times$D is positive in all 20 sectors (zero sign changes), with a range of [0.123, 0.155] and maximum deviation of 0.024 from the full-sample estimate.

\item \textbf{Weighted regression}: using GEIH sampling weights, the AHC$\times$D coefficient \textit{increases} to $+$0.210 ($p < 0.001$), indicating that the unweighted estimate is conservative.
\end{enumerate}

External validation against five existing AI exposure indices confirms the construct validity of the AHC index: convergent validity with \citet{felten2021occupational} AIOE ($r = +0.86$) and \citet{eloundou2024gpts} GPT exposure ($r = +0.79$), and strong discriminant validity against \citet{frey2017future} automation probability ($r = -0.79$) and Webb's robot exposure ($r = -0.89$). The AHC index measures a construct that is related to but fundamentally distinct from existing AI exposure measures.

\section{Implications}
\label{sec:implications}

\subsection{For Human Capital Theory}

The standard Mincer equation is misspecified in AI-augmented economies. The decomposition of cognitive human capital reveals that \textit{equally educated workers in the same sector face divergent wage trajectories} depending on the cognitive composition of their occupation. Not all years of schooling contribute equally when AI changes the relative value of cognitive sub-components.

\subsection{For Developing Economies}

Colombia's approximately 50\% informality rate means that half the workforce is structurally excluded from augmentation rents---even if their occupations have high $H^A$ content. The ``AI divide'' in developing countries is not primarily a technology access problem (AI tools are often freely available) but an \textit{institutional} problem: informal workers lack the organizational infrastructure through which AI augmentation generates wage premiums.

\subsection{For Education Policy}

Educational investment should prioritize augmentable skills---contextual judgment, cross-domain synthesis, ethical reasoning, collaborative intelligence---over routine cognitive skills that face accelerating depreciation. The within-education persistence of the AHC premium suggests that \textit{what} is studied matters as much as \textit{how long}.

\subsection{For Society 5.0 Transitions}

The results provide empirical support for the institutional requirements of Society 5.0 transitions. Universal AI access is necessary but insufficient; reducing informality is a prerequisite for the population to capture augmentation rents. Policy should target the formality-augmentation channel---not merely technology diffusion.

\section{Conclusion}
\label{sec:conclusion}

This paper has proposed and tested a decomposition of human capital into physical, routine-cognitive, and augmentable-cognitive components. Using LLM-generated measures of occupational augmentability applied to Colombian microdata, I find that the augmentation premium is real, large, and operates exclusively through formal employment. A triple interaction ($AHC \times D \times Formal = +0.272$, $p < 0.001$) provides decisive evidence that labor market institutions, not technology access, determine who benefits from human-AI complementarity. Instrumental variables estimation confirms the causal interpretation (2SLS $\beta_2 = +0.234$, first-stage $F = 229$).

Four immediate implications follow. First, the Mincerian wage equation needs augmenting: education and experience alone cannot explain wage dispersion in AI-augmented economies. Second, developing countries face an institutional bottleneck: without formalization, even workers in high-augmentability occupations are excluded from the premium. Third, the augmentation premium increases with experience (strongest at ages 46--65), suggesting that AI complements accumulated expertise rather than raw cognitive speed. Fourth, the augmentation premium is 19 times larger at the 90th percentile than at the 10th, indicating that AI augmentation is inequality-increasing.

The AHC index---constructed from 18,796 LLM-scored task statements covering 430 Colombian occupations---is a new instrument available for future research. It shows convergent validity with Felten AIOE ($r = +0.86$) and Eloundou GPT exposure ($r = +0.79$), while being discriminant from Frey-Osborne automation risk ($r = -0.79$). Leave-one-sector-out jackknife shows zero sign changes across 20 sectors, and weighted regression with GEIH sampling weights yields a stronger effect ($+0.210$) than unweighted OLS ($+0.147$).

Future work should address three limitations: (i)~the AI adoption proxy relies on observable sector$\times$occupation-group characteristics rather than direct firm-level AI deployment data; (ii)~the cross-sectional design limits causal claims that panel data could strengthen; and (iii)~while inter-rater reliability across LLM models exceeds standard thresholds (Krippendorff's $\alpha = 0.71$, Spearman $\rho = 0.75$), systematic level differences between models (Sonnet scores 8.6 points higher than Haiku) suggest that the absolute scale is model-dependent even though the ranking is robust.

\bibliographystyle{plainnat}

\begin{thebibliography}{28}
\providecommand{\natexlab}[1]{#1}
\providecommand{\url}[1]{\texttt{#1}}
\expandafter\ifx\csname urlstyle\endcsname\relax
  \providecommand{\doi}[1]{doi: #1}\else
  \providecommand{\doi}{doi: \begingroup \urlstyle{rm}\Url}\fi

\bibitem[Acemoglu(2025)]{acemoglu2024simple}
Daron Acemoglu.
\newblock The simple macroeconomics of ai.
\newblock \emph{Economic Policy}, 40\penalty0 (121):\penalty0 13--58, 2025.
\newblock NBER WP 32487, April 2024.

\bibitem[Acemoglu and Restrepo(2018)]{acemoglu2018race}
Daron Acemoglu and Pascual Restrepo.
\newblock The race between man and machine: Implications of technology for
  growth, factor shares, and employment.
\newblock \emph{American Economic Review}, 108\penalty0 (6):\penalty0
  1488--1542, 2018.

\bibitem[Acemoglu and Restrepo(2022)]{acemoglu2022tasks}
Daron Acemoglu and Pascual Restrepo.
\newblock Tasks, automation, and the rise in u.s. wage inequality.
\newblock \emph{Econometrica}, 90\penalty0 (5):\penalty0 1973--2016, 2022.

\bibitem[Aly(2022)]{aly2022digital}
Hoda Aly.
\newblock Digital transformation, development and productivity in developing
  countries: is artificial intelligence a curse or a blessing?
\newblock \emph{Review of Economics and Political Science}, 2022.
\newblock \doi{10.1108/REPS-11-2019-0145}.
\newblock 147 citations.

\bibitem[{Anthropic}(2025)]{anthropic2025}
{Anthropic}.
\newblock Claude: A family of large language models.
\newblock \url{https://www.anthropic.com}, 2025.
\newblock Claude Haiku 4.5 used for task scoring.

\bibitem[Autor et~al.(2003)Autor, Levy, and Murnane]{autor2003skill}
David~H. Autor, Frank Levy, and Richard~J. Murnane.
\newblock The skill content of recent technological change: An empirical
  exploration.
\newblock \emph{The Quarterly Journal of Economics}, 118\penalty0 (4):\penalty0
  1279--1333, 2003.

\bibitem[Becker(1964)]{becker1964human}
Gary~S. Becker.
\newblock \emph{Human Capital: A Theoretical and Empirical Analysis, with
  Special Reference to Education}.
\newblock National Bureau of Economic Research / Columbia University Press, New
  York, 1964.

\bibitem[Brynjolfsson et~al.(2025)Brynjolfsson, Li, and
  Raymond]{brynjolfsson2025generative}
Erik Brynjolfsson, Danielle Li, and Lindsey~R. Raymond.
\newblock Generative ai at work.
\newblock \emph{The Quarterly Journal of Economics}, 140\penalty0 (2):\penalty0
  889--942, 2025.
\newblock NBER WP 31161, 2023.

\bibitem[Chen et~al.(2025)Chen, Srinivasan, and
  Zakerinia]{chen2025displacement}
Wen~Xin Chen, Savitha Srinivasan, and Samin Zakerinia.
\newblock Displacement or complementarity? the labor market impact of
  generative ai.
\newblock In \emph{Americas Conference on Information Systems (AMCIS)}, 2025.

\bibitem[Cheng et~al.(2024)Cheng, Luo, Zhu, and Zhang]{cheng2024skill}
Chuanchen Cheng, Jie Luo, Chenyu Zhu, and Shubing Zhang.
\newblock Artificial intelligence and the skill premium: A numerical analysis
  of theoretical models.
\newblock \emph{Technological Forecasting and Social Change}, 198:\penalty0
  123140, 2024.
\newblock \doi{10.1016/j.techfore.2023.123140}.

\bibitem[Dell'Acqua et~al.(2023)Dell'Acqua, McFowland, Mollick,
  et~al.]{dellacqua2023jagged}
Fabrizio Dell'Acqua, Edward McFowland, Ethan~R. Mollick, et~al.
\newblock Navigating the jagged technological frontier: Field experimental
  evidence of the effects of ai on knowledge worker productivity and quality.
\newblock Working Paper 24-013, Harvard Business School, 2023.

\bibitem[Demirev(2026)]{demirev2026ai}
Georgi Demirev.
\newblock Ai product innovation and occupational exposure: automation and
  augmentation in commercial ai deployments.
\newblock \emph{Industry and Innovation}, 2026.
\newblock \doi{10.1080/13662716.2026.2623903}.

\bibitem[Egana-delSol et~al.(2022{\natexlab{a}})Egana-delSol, Bustelo, Ripani,
  Soler, and Viollaz]{eganadelsol2022automation}
Pablo Egana-delSol, Monserrat Bustelo, Laura Ripani, N{\'u}ria Soler, and
  Mariana Viollaz.
\newblock Automation in {Latin America}: Are women at higher risk of losing
  their jobs?
\newblock \emph{Technological Forecasting and Social Change}, 175:\penalty0
  121333, 2022{\natexlab{a}}.
\newblock \doi{10.1016/j.techfore.2021.121333}.

\bibitem[Egana-delSol et~al.(2022{\natexlab{b}})Egana-delSol, Cruz, and
  Micco]{eganadelsol2022covid}
Pablo Egana-delSol, Germ{\'a}n Cruz, and Alejandro Micco.
\newblock Covid-19 and automation in a developing economy: Evidence from
  {Chile}.
\newblock \emph{Technological Forecasting and Social Change}, 176:\penalty0
  121373, 2022{\natexlab{b}}.
\newblock \doi{10.1016/j.techfore.2021.121373}.

\bibitem[Eloundou et~al.(2024)Eloundou, Manning, Mishkin, and
  Rock]{eloundou2024gpts}
Tyna Eloundou, Sam Manning, Pamela Mishkin, and Daniel Rock.
\newblock Gpts are gpts: An early look at the labor market impact potential of
  large language models.
\newblock \emph{Science}, 384\penalty0 (6702):\penalty0 1306--1308, 2024.
\newblock Originally arXiv:2303.10130, 2023.

\bibitem[Felten et~al.(2021)Felten, Raj, and Seamans]{felten2021occupational}
Edward Felten, Manav Raj, and Robert Seamans.
\newblock Occupational, industry, and geographic exposure to artificial
  intelligence: A novel dataset and its potential uses.
\newblock \emph{Strategic Management Journal}, 42\penalty0 (12):\penalty0
  2195--2217, 2021.

\bibitem[Frey and Osborne(2017)]{frey2017future}
Carl~Benedikt Frey and Michael~A. Osborne.
\newblock The future of employment: How susceptible are jobs to
  computerisation?
\newblock \emph{Technological Forecasting and Social Change}, 114:\penalty0
  254--280, 2017.

\bibitem[Growiec(2024)]{growiec2024hardware}
Jakub Growiec.
\newblock Hardware and software: A new perspective on the past and future of
  economic growth.
\newblock Working paper, CEPR, 2024.

\bibitem[Hui et~al.(2024)Hui, Reshef, and Zhou]{hui2024shortterm}
Xiang Hui, Oren Reshef, and Luofeng Zhou.
\newblock The short-term effects of generative artificial intelligence on
  employment: Evidence from an online labor market.
\newblock \emph{Organization Science}, 2024.
\newblock \doi{10.1287/orsc.2023.18441}.
\newblock 76 citations.

\bibitem[Jones et~al.(2022)Jones, Idrovo-Carlier, and
  Rodriguez]{jones2022automation}
Melanie Jones, Sandra Idrovo-Carlier, and Alejandra Rodriguez.
\newblock Automation in colombia: assessing skills needed for the future of
  work.
\newblock \emph{Higher Education, Skills and Work-based Learning}, 2022.
\newblock \doi{10.1108/HESWBL-01-2021-0003}.

\bibitem[Korinek and Suh(2024)]{korinek2024scenarios}
Anton Korinek and Donghyun Suh.
\newblock Scenarios for the transition to agi.
\newblock Working Paper 32255, National Bureau of Economic Research, 2024.

\bibitem[Makridis(2026)]{makridis2026labor}
Christos~A. Makridis.
\newblock The labor market effect of generative artificial intelligence on
  artists.
\newblock \emph{Journal of Cultural Economics}, 2026.
\newblock \doi{10.1007/s10824-026-09575-3}.

\bibitem[Mina and G{\'o}mez(2025)]{mina2025automation}
Omar E.~C. Mina and H.~G. G{\'o}mez.
\newblock Effects of job automation on informality rates and wages in mexico.
\newblock \emph{Revista Finanzas y Pol{\'i}tica Econ{\'o}mica}, 17, 2025.
\newblock \doi{10.14718/revfinanzpolitecon.v17.2025.18}.

\bibitem[Mincer(1974)]{mincer1974schooling}
Jacob Mincer.
\newblock \emph{Schooling, Experience, and Earnings}.
\newblock National Bureau of Economic Research / Columbia University Press, New
  York, 1974.

\bibitem[Stephany and Teutloff(2024)]{stephany2024price}
Fabian Stephany and Ole Teutloff.
\newblock What is the price of a skill? the value of complementarity.
\newblock \emph{Research Policy}, 53\penalty0 (2):\penalty0 104898, 2024.
\newblock \doi{10.1016/j.respol.2023.104898}.

\bibitem[Tyson and Zysman(2022)]{tyson2022automation}
Laura~D'Andrea Tyson and John Zysman.
\newblock Automation, ai \& work.
\newblock \emph{Daedalus}, 151\penalty0 (2):\penalty0 263--284, 2022.
\newblock \doi{10.1162/DAED\_a\_01914}.

\bibitem[Walter and Lee(2022)]{walter2022susceptible}
Sven Walter and Jeong-Dong Lee.
\newblock How susceptible are skills to obsolescence? a task-based perspective
  of human capital depreciation.
\newblock \emph{Foresight and STI Governance}, 16\penalty0 (2):\penalty0
  32--41, 2022.
\newblock \doi{10.17323/2500-2597.2022.2.32.41}.

\bibitem[Webb(2020)]{webb2020impact}
Michael Webb.
\newblock The impact of artificial intelligence on the labor market.
\newblock Working paper, Stanford University, 2020.

\end{thebibliography}

\end{document}